\documentclass[aip,rsi,reprint,graphicx]{revtex4-1}
\usepackage{epsfig}
\usepackage{graphicx}
\usepackage{dcolumn}
\usepackage{bm}
\usepackage{braket}
\usepackage{amssymb}
\usepackage{amsmath}

\draft 

\begin{document}


\title{\textit{In-situ} Broadband Cryogenic Calibration for Two-port Superconducting Microwave Resonators} 

\author{Jen-Hao Yeh}
\email[]{davidyeh@umd.edu}
\affiliation{Electrical and Computer Engineering Department,
University of Maryland, College Park, Maryland 20742-3285}
\affiliation{CNAM, Physics Department, University of Maryland,
College Park, Maryland 20742-4111 }

\author{Steven M. Anlage}
\affiliation{Electrical and Computer Engineering Department,
University of Maryland, College Park, Maryland 20742-3285}
\affiliation{CNAM, Physics Department, University of Maryland,
College Park, Maryland 20742-4111 }


\date{\today}

\begin{abstract}
We introduce an improved microwave calibration method for use in a
cryogenic environment, based on a traditional three-standard
calibration, the Thru-Reflect-Line (TRL) calibration. The modified
calibration method takes advantage of additional information from
multiple measurements of an ensemble of realizations of a
superconducting resonator, as a new pseudo-Open standard, to correct
errors in the TRL calibration. We also demonstrate an experimental
realization of this \textit{in-situ} broadband cryogenic calibration
system utilizing cryogenic switches. All calibration measurements
are done in the same thermal cycle as the measurement of the
resonator (requiring only an additional 20 minutes), thus avoiding 4
additional thermal cycles for traditional TRL calibration (which
would require an additional 12 days). The experimental measurements
on a wave-chaotic microwave billiard verify that the new method
significantly improves the measured scattering matrix of a
high-quality-factor superconducting resonator.
\end{abstract}


\maketitle 

\section{Introduction}
Microwave resonators are widely-utilized devices in many science and
engineering fields. In order to achieve very low dissipation and
attain extremely high quality factors $Q$, microwave resonators can
be made of superconducting materials at temperatures lower than the
critical temperature $T_{c}$ \cite{Weinstock1999B, Hein1999B}. One
important application of superconducting resonators is in building
particle accelerators. They can transfer energy to a charged
particle beam with high efficiency because superconducting
resonators can store the energy with very low loss and a narrow
bandwidth \cite{Padamsee1998}.

Another application of superconducting resonators, and also the
motivation of the microwave calibration method in this paper, is to
test the predictions of wave chaos theories in a low loss (nearly
unitary) regime \cite{DarmstadtPapers}. Wave chaos, or quantum chaos
\cite{Stockmann1999B}, is a field where researchers study the
manifestations of chaotic dynamics of classical trajectories on the
short-wavelength (quantum or wave) properties of wave systems.
Random Matrix Theory (RMT) \cite{Mehta1991B} has successfully been
applied to predict many statistical properties of open wave-chaotic
systems, such as the scattering matrices ($S$), the impedance
matrices ($Z$), the conductivities, and the fading amplitudes
\cite{DarmstadtPapers, Brouwer1997, Fyodorov2005, Zheng2006,
Hemmady2005, Yeh2010, Yeh2012} in complicated wave scattering
environments. There has been great success in studying the
properties of wave chaotic systems in the context of microwave
billiards \cite{Stockmann1999B}.

For the scattering matrix $S$ of an open wave-scattering system, the
prediction of RMT is for the \textit{universal} features of the
statistical distributions of the elements of $S$ and their
correlations, which should only depend on the loss parameter of the
system \cite{Brouwer1997, Fyodorov2005, Zheng2006, Hemmady2005}. The
loss parameter $\alpha$ is defined as the ratio of the 3dB-frequency
bandwidth of the cavity resonances due to distributed losses, to the
average spacing between resonant frequencies, and it can be
expressed as $\alpha = k^{2}/(\triangle k^{2}_{n}Q)$, where $k$ is
the wave number for the incoming wave, $\triangle k^{2}_{n}$ is the
mean spacing of the adjacent eigenvalues of the Helmholtz operator
($\nabla^{2} + k^{2}$) of the corresponding closed system, and $Q$
is the loaded quality factor of the cavity \cite{Zheng2006,
Hemmady2005}. Examination of the predictions of RMT in very low loss
systems is interesting because extreme limits for the distribution
functions and other predictions are encountered \cite{Zheng2006,
Brouwer1997, Yeh2012}. For example, in a two-port time-reversal
invariant system, the distribution of the magnitude of $|S_{21}|$
can be approximated by a Rayleigh distribution when the system has a
moderate or high loss parameter, but the distribution $P(|S_{21}|)$
goes to a uniform distribution when the loss parameter goes to zero
\cite{Brouwer1997, Yeh2012}.

For a practical network system, the measured scattering matrix
contains both the universal features predicted by RMT and
system-specific features which are not included in the prediction of
RMT. The system-specific (\textit{non-universal}) features of the
wave scattering properties can be removed by methods such as the
random coupling model (RCM) \cite{Hemmady2005, Yeh2010, Hemmady2012,
Lawniczak2012} or the Poisson Kernel \cite{Mello1985}. In order to
test the statistical predictions, we use a two-port superconducting
microwave resonator as the wave-chaotic billiard to create a
low-loss experimental system, and we measure the $2\times 2$
scattering matrix $S$ of the superconducting cavity in a wide
frequency band. We then apply the random coupling model to reveal
the universal features from the measured scattering matrix to
compare with the predictions of RMT.

To measure the scattering matrix $S$ with high accuracy over a broad
bandwidth, a broadband calibration method of microwave measurement
is necessary. The calibration method removes the effect of the
transmission lines connecting the vector network analyzer (VNA) at
room temperature and the cryogenic sample at low temperature.
Because the convenient electronic-calibration kit of a commercial
VNA does not function in a cryogenic environment, one must perform a
manual calibration by utilizing known standards.

\section{Review of Cryogenic Microwave Calibration}
Microwave calibration is an important process to remove the
systematic errors due to the transmission lines and connectors
between the network analyzer and the device under test (DUT) as well
as other systematic measurement errors. The calibration process
utilizes measurements of known standards to move the reference plane
of the measurement to the ports of the DUT \cite{Agilent, Cano2010}.
A commonly-used calibration method for two-port measurement is the
Thru-Reflect-Line (TRL) calibration \cite{Engen1979, Marks1991}.

%
%

The Thru-Reflect-Line calibration uses three standards to calibrate
the effect of the two transmission lines connecting the two ports of
the DUT to the VNA. For the Thru standard measurement, the two
transmission lines are directly connected together; an additional
electrically short (on the order of one guided wavelength)
transmission line is added between the two transmission lines as the
Line standard; two identical reflectors are connected to the ends of
the two transmission lines as the Reflect standard \cite{Marks1991}.
The advantages of the TRL calibration are from two facts: (i) the
use of redundant calibration standards reduces the uncertainty due
to errors, such as connector irreproducibility, cable flexure,
test-set drift, and noise, and (ii) the foundation of the
calibration standard definitions depend solely on qualitative
requirements (uniformity of the lines, identical cross-sections of
the lines, and identical reflection coefficients of the Reflect
standards) \cite{Lewandowski2009}.


For applying TRL calibration in cryogenic systems, one challenge is
that measuring multiple standards may involve cooling down and
warming up of the system, changing the standard, and repeating the
thermal cycle. These thermal cycles can be very time consuming and
expensive. In addition, there is enhanced uncertainty of the
reproducibility of experimental conditions in different thermal
cycles \cite{Lewandowski2009, Cano2010}. On the other hand,
researchers have developed single-thermal-cycle calibration methods
which use an on-wafer cryogenic probe station or electromechanical
switches. However, the problem of these single-thermal-cycle methods
is that the differences between electrical paths in measurements of
different standards, which are assumed to be equal, degrade the
measurement accuracy and limit the frequency bandwidth
\cite{Cano2010}.

For one-port systems, a typical calibration method is the
Open-Short-Load (OSL) calibration. A pioneering work on broadband
cryogenic calibration of a one-port system was done by Booth
\textit{et al.} \cite{Booth1994} who measured the complex reflection
coefficient $S_{11}$ of a superconducting thin film. By taking
advantage of the superconductive feature, they cooled the sample
deep into the superconducting state and used it as a short-circuit
standard to correct their OSL calibration of the Corbino
reflectometer at room temperature (or temperatures higher than the
critical temperature $T_{c}$). Other researchers also applied the
OSL calibration and the superconducting-short-standard approach in
their one-port cryogenic measurement \cite{Stutzman2000,
Scheffler2005, Steinberg2012}. Instead of using the
superconducting-short-standard approach, Reuss and Richard
\cite{Reuss2000} also measured broadband $S_{11}$ of a
superconducting thin film, but they applied the 3-standard (Open,
Short, Load) calibration, once at room temperature and once at the
superconducting temperature (for a total of 6 calibration
measurements). One problem with the OSL calibration is that it is
very sensitive to the standards, and one must carefully reproduce
all connections, bending and twisting of the transmission lines, and
the temperature distribution in the apparatus \cite{Booth1994,
Reuss2000, Stutzman2000, Scheffler2005}. Kitano \textit{et al.}
\cite{Kitano2008} used the normal-state conductivity of a sample as
a Load standard to address this problem.

%
%
%
%
%

For two-port systems, researchers have developed cryogenic TRL
calibration methods for measurement of the $2\times 2$ scattering
matrix of a DUT. Laskar \textit{et al.} \cite{Laskar1996} utilized
the multiline method of the TRL and LRM (Line-Reflect-Match)
calibrations introduced by Marks \cite{Marks1991}, to their
cryogenic on-wafer probe station for noise and scattering-parameter
measurements, and they emphasized the importance of providing a
stable thermal environment \cite{Laskar1996}. Booth \textit{et al.}
also used a cryogenic probe station for scattering-parameter
measurements of their coplanar waveguide (CPW) structures in high
temperature superconductors \cite{Booth1997, Leong2006}. They
applied a set of CPW calibration structures of TRL standards to
characterize the errors in the network analyzer/probe station system
\cite{Booth1997}, or alternatively three other standards: a Thru, a
Reflect, and a series resistor \cite{Orloff2011}. Shemelin
\textit{et al.} used the TRL calibration for waveguides and coaxial
cables to measure ferrites at low temperature \cite{Shemelin2006}.
In addition to the TRL calibration method, Jun \textit{et al.} use a
different calibration method by introducing a cryogenic dip probe
for time-domain measurements of nanodevices \cite{Jun2004}.

%
%

Another way to achieve single-thermal-cycle TRL calibration is to
utilize cryogenic microwave switches \cite{Birx1979}. Ranzani
\textit{et al.} \cite{Ranzani2012} use cryogenic switches (coaxial
subminiature latching switches) to switch the coaxial cables from
the VNA to coaxial cables with different calibration standards, as
well as the device under test. The electromechanical switches simply
operate by means of brief electrical pulses to latch the switch to
different positions, so no electrical signal is applied to the
switch in its quiescent state. Due to the convenience of connecting
the coaxial transmission lines to the ports of our superconducting
cavity, we use cryogenic switches to develop the \textit{in-situ}
calibration system. Similar cryogenic switches have been applied for
calibrating measurement of Superconducting Quantum Interference
Device (SQUID) amplifiers \cite{Spietz2010}, measuring different
superconducting qubit samples \cite{Slichter2011}, or other
experiments involving superconducting quantum computing
\cite{MartinisWeb}.

%
%
%

Although the TRL calibration is less sensitive to the properties of
the standards than the OSL calibration, the TRL calibration is still
limited by errors in its assumptions, such as irreproducibility of
the transmission lines in each measurement, differences in the
reflection coefficients of the two reflectors, and irreproducibility
of the connector interface \cite{Lewandowski2009, Juroshek1987}.
Researchers have tried different methods to reduce the calibration
errors, such as the development of precise dimensional
characterization techniques for the transmission lines
\cite{MacKenzie1966}, the minimization of the possible
center-conductor-gap variation \cite{Hoffmann2007}, and modeling of
the electrical properties combined with self-calibration approaches
\cite{Vandersteen1997, Lewandowski2009}. In our cryogenic
measurements, the temperature dependence of the scattering matrices
of all transmission lines and imperfect TRL standards become
additional sources of errors. These small errors are especially
significant in our extremely-low loss system because the calibrated
$|S_{11}|$ and $|S_{22}|$ are very close to 1 away from the
resonance frequencies.

In this paper, we introduce a self-calibration approach by taking
advantage of multiple measurements in different realizations of the
superconducting resonator. We call this additional information a
pseudo-Open standard and utilize it to correct the measured
scattering matrices after the TRL calibration. With the improvement
provided by the pseudo-Open standard, we also demonstrate an
experimental result utilizing this \textit{in-situ} broadband
cryogenic calibration system.

\section{Calibration Method}
\subsection{The Superconducting Resonator}

\begin{figure}
\includegraphics[width=2.8in]{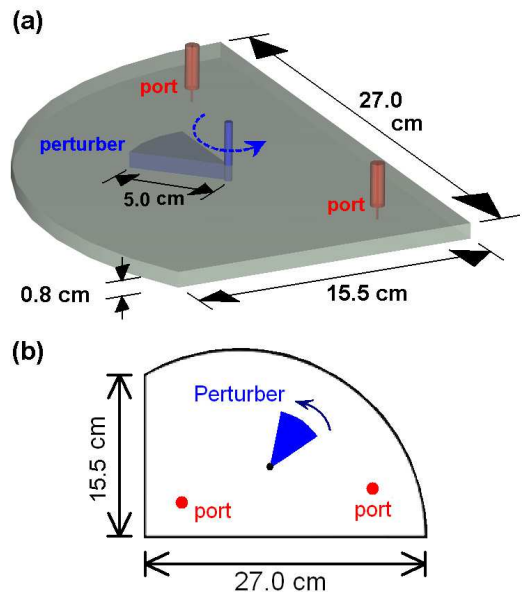}
\caption{(a) The quasi-two-dimensional cut-circle microwave cavity
in a three-dimensional perspective, showing the cavity dimensions,
ports, and the perturber. (b) Projected two-dimensional view of the
cut-circle billiard.}\label{cutcircle}
\end{figure}

We have carried out experiments by measuring the complex $2\times2$
scattering matrix $S$ of a quasi-two-dimensional microwave cavity,
illustrated in Fig. \ref{cutcircle}. There are two coupling ports
(the red cylinders and dots in Fig. \ref{cutcircle}) where
microwaves are injected through an antenna attached to a single-mode
coaxial transmission line of characteristic impedance $Z_{0} = 50$
 $\Omega$. Each antenna is inserted into the cavity through a small
hole in the lid, similar to previous setups \cite{Hemmady2005,
Yeh2012}. The waves introduced have frequencies from 3 to 18 GHz,
and they are quasi-two-dimensional due to the thin height of the
cavity (0.8 cm in the $z$ direction). The antennas are terminated
with SMA connectors, one male and the other female, on the surface
of the resonator. The shape of the cavity is a ``cut-circle'' and is
a billiard potential that shows classical chaos \cite{Yeh2012,
Ree1999, Dietz2006, Dietz2008}.

The superconducting cavity is made of copper with Pb-plated walls
and cooled to a temperature (6.6 K) below the transition temperature
of Pb \cite{Dietz2006, Dietz2008, Richter2001}. Measurements of the
transmission spectrum suggest that the quality factor of the
resonances is on the order of $Q \approx 10^{5}$. A Teflon wedge
(the blue wedge in Fig. \ref{cutcircle}) can be rotated as a
ray-splitting perturber inside the cavity, and we rotate the wedge
by $5^{o}$ each time to create an ensemble of 72 different
realizations. The rotation axis of the perturber is connected with a
MDC vacuum rotary feedthrough (BRM-133) at room temperature, so we
can control the angular position from outside the cryostat. The
original purpose of creating these realizations was to gather
statistics to test the statistical predictions of RMT, but it is now
a critical procedure for the pseudo-Open standard.

\subsection{The \textit{In-situ} Broadband Cryogenic
Calibration System}

\begin{figure}
  \includegraphics[width=3in]{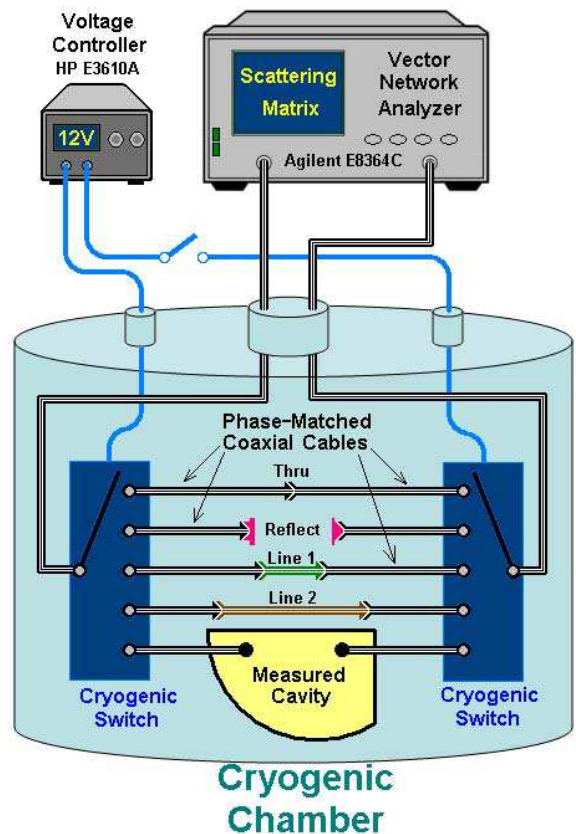}
  \caption{Schematic experimental setup of the \textit{in-situ} broadband cryogenic
calibration system. The 5 pairs of phase-matched coaxial cables have
nearly identical length and are not shown to
scale.}\label{cryoSetup}
\end{figure}

Figure \ref{cryoSetup} shows the setup of our \textit{in-situ}
broadband cryogenic calibration system. The term
``\textit{in-situ}'' means the TRL calibration process can be
applied at low temperatures without spending a great deal of time
changing standards. Here we utilize two cryogenic 6-position
switches (Radiall coaxial subminiature latching switches R591722605)
to include one Thru, one Reflect, and two Line standards together
with the cavity under measurement. Each switch is connected to
different standards, or to the cavity, by RF COAX phase-matched (13
inch long between interfaces, electrical length deviations $<$ 1 ps)
SMA coaxial cables (S086MMHF-013-1). These cables have male SMA
connectors in the both ends. Therefore, for the Thru standard, we
connect a pair of the coaxial cables with a female-female adapter
(Mini-Circuits adapter SF-SF50+). In order to make sure the
electrical paths are as identical as possible, the same type of
industrially-assembled cables, and adapters, have been used for the
other 4 pairs of transmission lines. In future experiments, we plan
to replace half of the phase-matched coaxial cables by cables
terminated with one male connector and one female connector. In this
way we will not need the female-female adapters and can reduce the
number of connectors in each electrical path, so the number of error
sources can also be reduced. For the Reflect standard, we use two
short circuits, one with a male connector and the other with a
female connector (Fairview Microwave models SC2136 and SC2141), to
terminate the pair of coaxial cables. For the Line standards, we add
two different SMA male-to-female adapters of electrical length 1.94
cm and 2.56 cm (Fairview Microwave models SM4971 and SM5291) to
connect the two pairs of coaxial cables. All of these standards and
the cut-circle cavity are at a uniform temperature in the cryostat,
and the switches are controlled by voltage pulses from a DC power
supply (Hewlett-Packard E3610A) outside the cryostat. The network
analyzer is an Agilent Technologies E8364C.

The cryostat includes two cylindrical vacuum chambers (aluminum)
with one cylindrical thermal shield (copper) in between them. The
switches, calibration standards, and the superconducting resonator
are all contained in the inner vacuum chamber (designed temperature
4 K, radius 16.5 cm, and height 30.2 cm). We use an Alcatel Drytel
31 Dry Vacuum Pump System to evacuate the inner and outer chambers
to a pressure lower than $1\times10^{-6}$ atm and a Cryomech PT405
Pulse Tube Cryorefrigerator (with a water-cooled compressor) to cool
down the resonator to the base temperature of 6.6 K. The resonator
is hung on the cold plate of the Cryorefrigerator, and we also use
copper thermal straps to connect the cold plate and the resonator.
The thermometer is attached in the lower part of the outside surface
of the resonator. We also designed two copper clamps to mount and
thermally anchor the cryogenic switches on the cold plate. The
second layer of the cryostat is the cylindrical copper shield
(designed temperature 40 K, radius 17.8 cm, and height 43.8 cm). The
third layer is the outer vacuum chamber (designed temperature 300 K,
radius 20.3 cm, and height 55.9 cm). For a thermal cycle, it takes
about one day to pump the system to vacuum and cool down the
resonator to thermal equilibrium at the base temperature, while
warming-up takes about two days. The dwell time of the system at the
base temperature can be longer than one week.

One advantage of using cryogenic switches is to save a great deal of
time for the TRL calibration. One full thermal cycle of this
cryostat takes about 3 days, so a multiple-thermal-cycle TRL
calibration for 4 standards would require an additional 12 days for
calibration. However, with the \textit{in-situ} calibration system,
we only need an additional 20 minutes for measuring the 4 standards
within the same thermal cycle for measuring the superconducting
cavity. The cryogenic switches also make the calibration process
conveniently accessible each time the system environment is
modified, for example, by changing the temperature, microwave power,
or external magnetic field. We can adjust the equilibrium
temperature of the cavity and standards by tuning the applied DC
current to heating resistors installed in the cryostat. The applied
microwave power can be controlled by the network analyzer. In
addition to increasing the efficiency of the experiment, avoiding
opening/closing the chamber for changing standards also reduces the
uncertainties created by having different temperatures or different
layouts of transmission lines in each measurement.

On the other hand, the disadvantage of using cryogenic switches is
that the switches utilize 5 different pairs of transmission lines to
connect to the standards or the cavity. The differences of the
scattering matrices of these electrical paths are additional errors
of the TRL calibration. We have measured the differences at room
temperature, and the deviations from the switches are $|\triangle S|
< 0.01$; the deviations from the transmission lines are $|\triangle
S| < 0.06$. These errors can be reduced from the calibrated results
by the pseudo-Open standard introduced in the following section.

The calibration system is also broadband because we install two Line
standards. The TRL calibration is invalid for frequencies where the
phase difference between the Thru standard and the Line standard is
too small (the phase difference should be greater than 20 degrees
and less than 160 degrees)\cite{Agilent, Marks1991}. To solve this
problem, one can use two Line standards with different lengths and
make sure that the problematic frequency bands do not overlap. In
our experiment, one Line standard has problematic frequency bands
near 7.7 GHz and 15.3 GHz; the other Line standard has problematic
frequency bands near 5.9 GHz, 11.8 GHz, and 17.6 GHz. Therefore, we
use the cryogenic switches to connect two different Line standards
to achieve broadband calibration, for example, we can measure the
scattering matrix continuously from 3 to 18 GHz.

\begin{figure}
\includegraphics[width=3.4in]{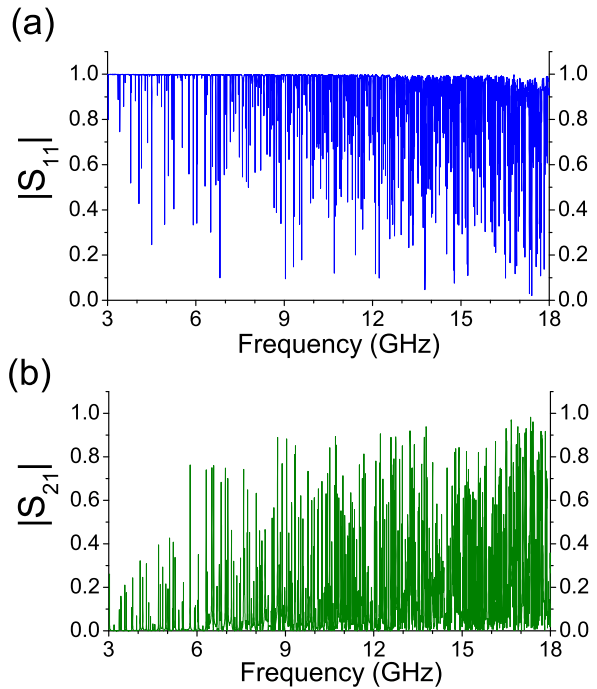}
\caption{The magnitude of (a) $|S_{11}|$ and (b) $|S_{21}|$ with
high-quality-factor resonances from 3 to 18 GHz of a single
realization of the superconducting cavity in the \textit{in-situ}
cryogenic calibration system. The data have been calibrated with the
TRL calibration and corrected by the pseudo-Open
method.}\label{broadband}
\end{figure}

We use MATLAB to operate the TRL calibration according to Rytting's
algorithm \cite{RyttingWeb}, and we combine the \textit{good}
frequency bands from the two Line standards to create a broadband
result. We then use the pseudo-Open standard to remove the remaining
errors in the TRL-calibrated data. One example of the result after
these procedures is shown in Fig. \ref{broadband} as the $|S_{11}|$
and $|S_{21}|$ of a single realization from 3 to 18 GHz. We can see
the resonance density increase with frequency. In low frequency
regions, the resonances are sharp and well-separated, and $|S_{11}|$
is close to 1 in the frequencies away from the resonant frequencies.
In high frequency regions, the resonance density increases, and the
resonances start to overlap with each other. For example, we do not
see the off-resonance background close to 1 in $|S_{11}|$ near 18
GHz.

\subsection{Pseudo-Open Standard}

The scattering matrices calibrated by the TRL method still have many
errors due to the fact that the practical standards and coaxial
cables do not perfectly satisfy the assumptions of the TRL
calibration. The remaining issues are: (1) the irreproducibility of
the transmission lines and connectors in each electrical path, (2)
the difference between the two reflectors in the Reflect standard,
and (3) the impedance-mismatch and the imperfection of the Thru and
Line standards. These errors are especially critical when $|S_{11}|$
and $|S_{22}|$ of the measured cavity are close to 1. In our case,
while the frequency is away from the resonant frequencies of the
superconducting resonator, $|S_{11}|$ and $|S_{22}|$ are very close
to 1 (i.e. the transmission coefficient $|S_{21}|$ is very close to
0, and there is almost no absorption in the cavity) in the extremely
low loss environment. Therefore, a small error can make $|S_{11}|$
or $|S_{22}|$ larger than 1 and cause non-physical results. This
small error is also critical for analysis of the $2\times 2$
impedance matrix $Z$, obtained by a bilinear transformation as $Z =
Z_{0} (I_{2}+S)/(I_{2}-S)$ where $Z_{0}$ is the diagonal
characteristic impedance matrix of the transmission lines
($50\Omega$) and $I_{2}$ is a $2\times 2$ identity matrix. The
denominator $(I_{2}-S)$ makes the impedance matrix sensitive to this
small error when $S_{11}$ or $S_{22}$ are close to 1.

\begin{figure}
\includegraphics[width=3in]{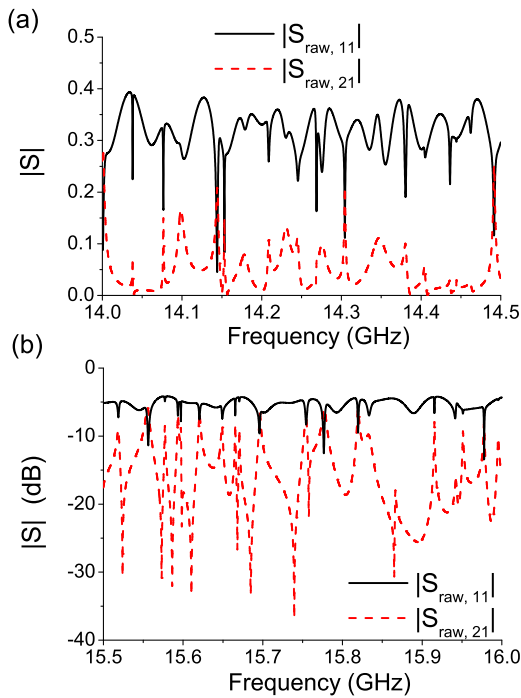}
\caption{The magnitude of raw measured scattering matrix element
$S_{11}$ (black solid curves) and $S_{21}$ (red dash curves) versus
frequency for the superconducting cut-circle microwave cavity at 6.6
K. (a) An example with the frequency band from 14.0 to 14.5 GHz
plotted in linear scale; (b) another example with the frequency band
from 15.5 to 16.0 GHz plotted in semi-logarithmic
scale.}\label{Sraw}
\end{figure}

\begin{figure}
\includegraphics[width=3in]{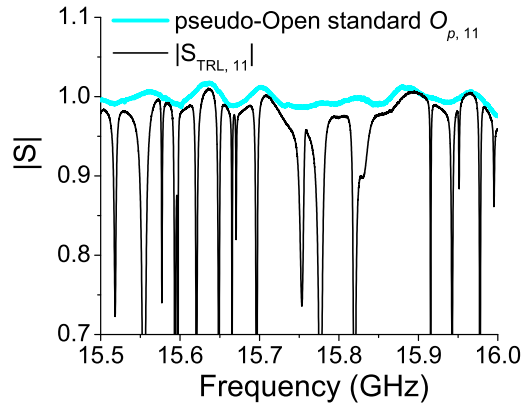}
\caption{The magnitude of TRL-calibrated $S_{11}$ (the black curve)
and the pseudo-Open standard (the thicker light-blue curve) versus
frequency.}\label{Scali}
\end{figure}

Figure \ref{Sraw} shows two examples of the raw data (without
calibration) of the magnitude of the scattering matrix elements
versus frequency. Because the loss parameter in the superconducting
resonator is very low ($\alpha \ll 1$, i.e. high $Q$), the curves of
$|S_{raw, 11}|$ and $|S_{raw, 21}|$ show sharp and well-separated
resonances on a smoothly varying background. Note that all
degeneracies are broken in wave-chaotic billiards
\cite{Stockmann1999B}. The background feature shows the influence of
the transmission lines between the cavity and the network analyzer,
and this is what we want to remove by calibration. With the
cryogenic TRL calibration, we can eliminate most of the influence
from the transmission lines. One example of the TRL-calibrated
result $|S_{TRL, 11}|$ of a single realization is shown as the black
curve in Fig. \ref{Scali}. As we expect for a superconducting
resonator, now $|S_{TRL, 11}|$ is close to 1 at frequencies away
from the resonant frequencies. However, there are still small
variations in the background due to the systematic errors of the TRL
calibration. Note that at some frequencies (e.g. near 15.64 GHz),
the small error makes $|S_{TRL, 11}| > 1$.

\begin{figure}
\includegraphics[width=3in]{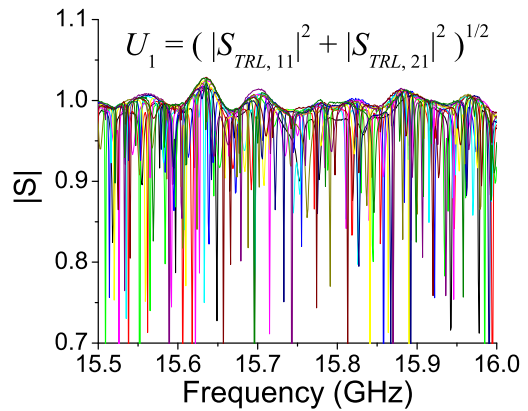}
\caption{Shown are 12 examples of different realizations of
$U_{1}=(|S_{TRL, 11}|^2+|S_{TRL, 21}|^2)^{1/2}$ in varied colors
versus frequency.}\label{allSpo}
\end{figure}

To solve this problem, we introduce the pseudo-Open standard by
taking advantage of the multiple measurements of the 72 ensemble
realizations of the cavity. In each realization, we only change the
orientation of the Teflon perturber (see Fig. \ref{cutcircle}), and
all of the other features, including the transmission lines, cavity
volume, coupling, etc., remain the same. Therefore, by comparing the
measured data of the 72 realizations, we see a
systematically-varying background in the scattering matrices in all
realizations, but the narrow and well-separated resonances move to
various frequencies. For illustration, Fig. \ref{allSpo} shows the
quantity $(|S_{TRL, 11}|^2+|S_{TRL, 21}|^2)^{1/2}$ in 12
realizations (each for a perturber orientation $30^{o}$ apart). Note
that the resonances occur at varied frequencies in different
realizations, but the off-resonance regions form a
systematically-varying background.

We note that $U_{1}\equiv(|S_{TRL, 11}|^2+|S_{TRL, 21}|^2)^{1/2}$
and $U_{2}\equiv(|S_{TRL, 22}|^2+|S_{TRL, 12}|^2)^{1/2}$ are unity
in a lossless system. For a very low loss system, RMT predicts that
the $U_{1}$ and $U_{2}$ of an ensemble of realizations have
statistical distributions \cite{Brouwer1997} where most of the
samples are close to 1. If the cavity is weakly coupled through the
ports to outside (i.e. $|S_{21}|$ and $|S_{12}|$ are closer to 0;
$|S_{11}|$ and $|S_{22}|$ are closer to 1), the distributions show
that the samples of $U_{1}$ or $U_{2}$ are even closer to 1.
Therefore, if one takes the maximum of many realizations of $U_{1}$
or $U_{2}$ under these conditions, it should be very close to 1.

We have constructed a numerical model to analyze the reason for the
systematically-varying background. We use RMT and the RCM to
numerically generate multiple-realization data to represent a
superconducting cavity, and the maxima over the realization ensemble
of $U_{1}$ and $U_{2}$ are very close to 1 for every frequency.
However, if we combine these numerical cavity data with the measured
data of our TRL standards and coaxial cables, and carrying out the
same TRL calibration, we see a similar systematically-varying
background. Therefore, this frequency-dependent feature represents a
combination of all errors of the TRL calibration. According to our
model, the major error in our experiment is from the Line standards,
and it can cause errors for $S_{11}$ and $S_{22}$ of $|\triangle
S|<0.1$ and an error for $S_{21}$ of $|\triangle S|<0.04$. The
errors from the Thru standard and the Reflect standard are all
smaller by a factor of 2 to 3.

In order to utilize this feature, we take the maximum values of
$U_{1}$ and $U_{2}$ of the experimental data over the 72
realizations, and we define a frequency-dependent diagonal matrix
$O_{p}$, where $O_{p,11}$ (or $O_{p,22}$) is the maximum of $U_{1}$
(or $U_{2}$) over the 72 realizations. In a very low loss system or
a weakly coupled system, $O_{p}$ should be close to the identity
matrix if the TRL calibration has no error. Thus, the
frequency-dependent variations of $O_{p}$ represent the remaining
errors in the experimental data after the TRL calibration. We call
$O_{p}$ the pseudo-Open standard because $O_{p}$ is like an Open
standard when we exclude all resonances (i.e. no energy is
transmitted through the two ports). Figure \ref{Scali} shows the
pseudo-Open standard response $O_{p, 11}$ as the thicker light-blue
curve. By utilizing the information from multiple measurements of
the cavity in different realizations, the pseudo-Open standard helps
to calibrate out the errors due to the deviations between the
transmission lines connected to the TRL standards and the
transmission lines connected to the cavity.

\begin{figure}
\includegraphics[width=3in]{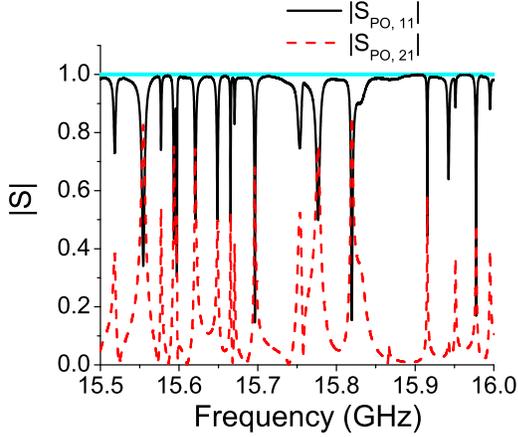}
\caption{ The magnitude of TRL-calibrated and pseudo-Open-corrected
$S_{11}$ (the black solid curve) and $S_{21}$ (the red dash curve)
versus frequency. The horizontal blue line shows
$|S|=1$.}\label{Spo}
\end{figure}

We simply remove the errors and obtain a TRL-calibrated and
pseudo-Open-corrected scattering matrix $S_{PO}$ by
\begin{equation}\label{modifiedS}
    S_{PO}=S_{TRL}O_{p}^{-1}.
\end{equation}
Figure \ref{Spo} shows the result of the pseudo-Open correction. The
small systematic variations in the TRL-calibrated data
$|S_{TRL,11}|$ is removed after applying the pseudo-Open standard.
Therefore, by using the TRL calibration with the pseudo-Open
standard, we can have well-calibrated data of the scattering matrix,
and we are able to do further analysis of the statistics of the
scattering matrices or the impedance matrices for wave chaos
research \cite{Yeh2012}.

We also use our numerical model to test how the pseudo-Open method
resolves the errors from the TRL calibration. Since the pseudo-Open
standard is based on the maximum values of $(|S_{TRL,
11}|^2+|S_{TRL, 21}|^2)^{1/2}$ and $(|S_{TRL, 22}|^2+|S_{TRL,
12}|^2)^{1/2}$, it contains no information about the phase of the
elements of the scattering matrix. Therefore, it can only correct
the magnitude of the scattering matrix, and the improvement is
better when $|S_{11}|$ and $|S_{22}|$ are closer to 1. Other
limitations concern determination of the maxima of $(|S_{TRL,
11}|^2+|S_{TRL, 21}|^2)^{1/2}$ and $(|S_{TRL, 22}|^2+|S_{TRL,
12}|^2)^{1/2}$. In order to get a sampled maximum value close to the
true maximum value, one needs to have a very low loss system or a
weakly coupled system, or one needs to have a large number of
realizations. Our experiment satisfies these requirements for the
pseudo-Open standard, and according to the numerical test, the
pseudo-Open method can remove the errors of the TRL calibration as
measured by the statistical distributions of the magnitudes of the
scattering matrix.

\subsection{Results and Test of RMT Predictions}

\begin{figure}
\includegraphics[width=3in]{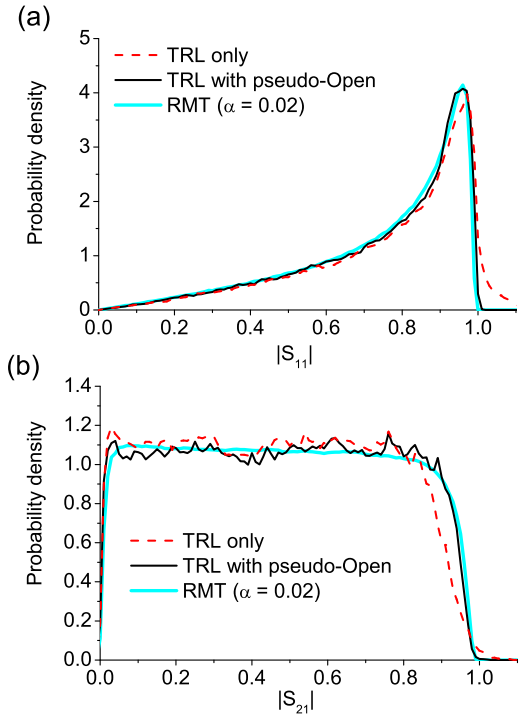}
\caption{The probability density of (a) $|S_{11}|$ and (b)
$|S_{21}|$ of the RMT predictions (thicker light-blue curves), the
experimental data from the superconducting cut-circle cavity at 6.6
K with TRL-calibration (red dash curves), and the data with
TRL-calibration and pseudo-Open correction (black
curves).}\label{Spdf}
\end{figure}

With the well-calibrated data of the scattering matrix, we can now
apply the random coupling model \cite{Hemmady2005, Zheng2006,
Yeh2010, Hemmady2012} to remove the system-specific features of the
scattering matrix and reveal the universal statistics which are
predicted by Random Matrix Theory \cite{Brouwer1997, Yeh2012}.
Figure \ref{Spdf} shows a comparison of the universal statistics
predicted by RMT (thicker light-blue curves) and the experimental
data which are calibrated with cryogenic TRL calibration only (red
dash curves) and also with pseudo-Open correction (black curves) in
terms of the distributions of the RCM-normalized $|S_{11}|$ and
$|S_{21}|$. The probability density functions (PDFs) of the
experimental data are taken from all 72 realizations and in
frequency from 14.0 to 16.0 GHz. The RMT predictions are the
best-matched PDFs with a single parameter (the loss parameter
$\alpha$), and $\alpha = 0.02$ is the fit value. The results show
that the pseudo-Open correction makes significant improvement in the
PDFs when $|S_{11}|$ or $|S_{21}|$ are close to 1. The non-physical
features ($|S|>1$, seen in the TRL calibrated data in Fig.
\ref{Scali}) are almost entirely eliminated with the correction of
the pseudo-Open standard. The PDFs of the experimental data are not
as smooth as the theoretical PDFs because we generate many more
numerical samples to plot the theoretical curves. One can generalize
these results to other resonant systems in which the modes can be
perturbed by external means, such as strain, electric field,
temperature, magnetic field, etc, to generate an ensemble of
multiple measurements for the pseudo-Open method.

\section{CONCLUSION}
Superconducting microwave resonators are useful devices in many
applications, and well-calibrated measurement of their scattering
matrices is important. In this paper we demonstrate an
\textit{in-situ} broadband cryogenic calibration system where the
calibration process is made dramatically more convenient by
installing two cryogenic switches for single-thermal-cycle TRL
calibration. We also introduce a pseudo-Open standard by taking
advantage of the ensemble realizations of the superconducting cavity
with a movable perturber and the feature of well-separated
resonances in an extremely low loss environment. Experimental data
verify that the pseudo-Open standard can significantly improve the
TRL-calibrated data. We show that the well-calibrated scattering
matrices are beneficial for wave chaos research, and this work
should broadly benefit various applications related to
high-precision cryogenic measurement.

\begin{acknowledgments}
We thank E. Ott and T. M. Antonsen for their helpful suggestions,
the group of A. Richter (Technical University of Darmstadt) for
graciously loaning the cut-circle billiard, and H. J. Paik and M. V.
Moody for use of the pulsed tube refrigerator. We also acknowledge
L. Ranzani for introducing us to the cryogenic switches. This work
is funded by the ONR/Maryland AppEl Center Task A2 (Contract No.
N000140911190), the AFOSR under Grant No. FA95500710049, NSF-GOALI
ECCS-1158644, and the Center for Nanophysics and Advanced Materials
(CNAM).
\end{acknowledgments}


\end{document}